\documentclass[letterpaper]{article}

\usepackage[T1]{fontenc}

\usepackage{geometry}
\usepackage{setspace}

\usepackage[style = chem-acs, articletitle = true]{biblatex}
\usepackage{amsmath}
\usepackage{textcomp}
\usepackage{gensymb}
\usepackage{hyperref}
\usepackage{graphicx}
\usepackage{float}
\newfloat{scheme}{htbp}{los}
\floatname{scheme}{Scheme}
\floatname{chart}{Chart}
\newfloat{graph}{htbp}{loh}

\usepackage{authblk}
\author[1]{Sergey A. Shteingolts}
\author[1]{Salman N. Salman}
\author[2]{Ron Levie}
\author[1]{Dan Mendels*}
\affil[1]{The Wolfson Department of Chemical Engineering, Technion -- Israel Institute of Technology, Haifa, 32000, Israel}
\affil[2]{Faculty of Mathematics, Technion -- Israel Institute of Technology, Haifa 32000, Israel}
\date{*Email: danmendels@technion.ac.il}

\title{Out-of-Distribution Inverse Design of Elastic Networks with Differentiable Graph Neural Network Molecular Dynamics}
\begin{document}

\maketitle

\begin{abstract}
Machine-learning-based inverse design can accelerate the discovery of materials with targeted properties, but conventional structure--property models often require large training datasets and generalize poorly beyond their training distribution. Here, we present a differentiable inverse design framework based on a graph neural network molecular dynamics simulator. By combining a short dynamical initialization with physics-based refinement during simulation, the framework enables optimization well beyond the conditions represented in the training data. Using disordered elastic networks, we show that a simulator trained only on non-auxetic systems with Poisson's ratios between $0.1$ and $0.4$ can design strongly auxetic networks with values as low as $-0.3$. The framework also produces stress--strain responses outside the range observed during training, creates localized mechanical defects absent from the training data, and generalizes across system size, enabling optimization of networks tested up to 5000 nodes despite being trained only on networks with fewer than 200 nodes.
\end{abstract}

\section{Introduction}

Traditional computational molecular and materials discovery largely follows a forward design paradigm, in which candidate structures or configurations are iteratively evaluated against target objectives. Because each evaluation often requires costly physics-based calculations, this process is computationally intensive and fundamentally limits the exploration of vast design spaces.\cite{Cheng2025AI-drivenMini-review, Sherman2020} Inverse design offers a compelling alternative by searching directly for physical configurations that exhibit predefined properties.\cite{Lin2026MachineApplications, Zunger2018InverseFunctionalities} Modern approaches increasingly employ machine-learning (ML) structure--property models, including generative architectures, as fast surrogates for physics-based calculations. Once trained, these models can evaluate properties orders of magnitude faster, enabling large-scale screening in the forward direction and making inverse design practical through efficient search, optimization, and generation of structures with targeted properties.\cite{Chen2021GenerativeMaterials, Lu2022InverseDiscovery, Wines2023InverseModels} However, such approaches tend to require large training datasets and struggle significantly with out-of-distribution (OOD) generalization, which can be essential for discovering novel engineering solutions.\cite{Mendels2022a, Mendels2022, Nandy2022AudacityDiscovery, Dou2023MachineScience, Hu2024RealisticLearning, Quesada-Molina2025NavigatingMaterials, Tenorio2026Out-of-distributionOpportunities, Smit2026TheDiscovery, Wu2024CompositionalDesign, Li2024Out-of-distributionFine-tuning} 

One proposed strategy for overcoming these hurdles draws on the concept of ML-learned collective variables \cite{Mendels2018,Mendels2018a,Piccini2018,Rizzi2019} developed in the context of enhanced sampling methods. In this framework, machine-learning models are used to learn the mapping between a system's elementary components and interactions and its functional modes, thereby enabling the identification of the salient microscopic features associated with these modes and their targeted modification to achieve desired functional changes \cite{Mendels2022a, Mendels2022, Medaparambath2026CollectivePeptide, Zhilkin2026GuidingBarriers, zick2026gradientbasedinversedesignfreeenergy}. 

Alternatively, it has recently been proposed that more data-efficient ML frameworks with improved OOD generalization can be developed by training models directly on dynamical data obtained from simulations or experiments.  Rather than learning static structure-to-property mappings, these models learn the underlying dynamics and can subsequently be used to evaluate physical properties from the resulting trajectories.\cite{Salman2025EvaluatingNetworks} One way to realize this approach is to train autoregressive ML-based simulators that predict a system's next configuration from its current state and iteratively generate full trajectories, thereby learning the underlying patterns that govern the dynamics of a given class of systems. When embedded within an optimization framework, such simulators can be used to iteratively modify an initial system configuration to achieve desired dynamical behaviors and macroscopic properties.\cite{Allen2022b}

A critical challenge in implementing such a pipeline for gradient-based inverse design is that the ML-based simulator must generate accurate trajectories from structural information alone, i.e., under structure-only initialization (SOI). In practice, however, accurate dynamical predictions often require additional state information, such as particle velocities and accelerations, and may also depend on a history of preceding configurations. Such information is generally unavailable in inverse design settings, where each simulation must be initialized from a newly proposed static configuration. \cite{Allen2022b, Yang2022LearningNetworks, Salman2025EvaluatingNetworks, Shteingolts2026EnablingSimulators} To address this limitation, recent work has proposed two SOI strategies: a simulator-cascade approach, in which the initial rollout steps are handled by separately trained neural networks, and a bootstrap approach, in which a short transient trajectory is generated by a custom differentiable MD engine. Both approaches provide the dynamical context required to initialize the ML-based simulator within the inverse design framework.\cite{Shteingolts2026EnablingSimulators} 

Another potential challenge in applying an ML-based simulator within an inverse design framework is ensuring reliable performance outside the training distribution. This is particularly important for inverse design, where the objective is to identify novel configurations with targeted functionalities. Although dynamical ML-based simulators have been shown to outperform static structure--property models in OOD settings, their generalization remains limited when the system dynamics change substantially.\cite{Salman2025EvaluatingNetworks} To address this challenge and extend the OOD capabilities of ML-based simulators, we recently introduced the Inference-Time Physics-Based Optimization (ITPO) approach.\cite{Shteingolts2026EnablingSimulators} ITPO ensures that the system satisfies predefined physics-based constraints during the rollout by refining the ML-based simulator's initial predictions at regular intervals. It has been shown to enable high-quality dynamical predictions even in regimes where the system exhibits qualitatively different and more complex dynamics than those present in the training data.

In this work, we present a novel framework for OOD inverse design based on an autoregressive, differentiable graph neural network molecular dynamics simulator (GMDS). Previously, we showed that an ML-based MD simulator that relied solely on static structural input performed poorly and thus could not be used in the optimization framework.\cite{Shteingolts2026EnablingSimulators} However, when the simulator is initialized using minimal dynamical information generated by a custom differentiable bootstrap MD simulator and combined with the ITPO scheme during prediction rollouts, the framework enables substantial OOD optimization. The resulting systems exhibit macroscopic functional behavior substantially different from that represented in the training data, together with qualitatively different and more complex dynamics.

We investigate the proposed methodology in the context of disordered elastic networks (DENs), highly versatile model systems whose structures can be precisely tuned to realize a wide range of mechanical and thermodynamic behaviors.\cite{Bahar2010, Reid2018, Reid2019, Rens2019, Mendels2022, Shen2024, Zu2025}  Specifically, we consider three design objectives: auxetic optimization, local defect embedding, and stress--strain curve tuning. Auxetic materials are uniquely characterized by a negative Poisson's ratio $\nu=-\frac{dy}{dx}$, meaning they contract transversely rather than expand when compressed longitudinally. For auxetic optimization, we demonstrate that the proposed inverse design framework, incorporating a GMDS trained exclusively on configurations with relatively high Poisson's ratios, $\nu \in [0.1,0.4]$, can generate strongly auxetic DENs with Poisson's ratios as low as $-0.3$. We further show that the same framework can engineer localized auxetic pockets within globally non-auxetic networks and concentrate a larger fraction of the network stress within prescribed local regions without substantially altering the macroscopic stress, despite the absence of local inhomogeneities or defects in the training data. In addition, the framework can optimize DEN stress--strain responses beyond the range represented in the training distribution. Finally, we demonstrate effective generalization across system sizes, enabling the optimization of DENs with up to 5,000 nodes despite the GMDS having been trained exclusively on networks containing approximately 200 nodes.

\section{Methods}

\subsection{Optimization algorithm}

\begin{figure}[ht]
  \centering
  \includegraphics{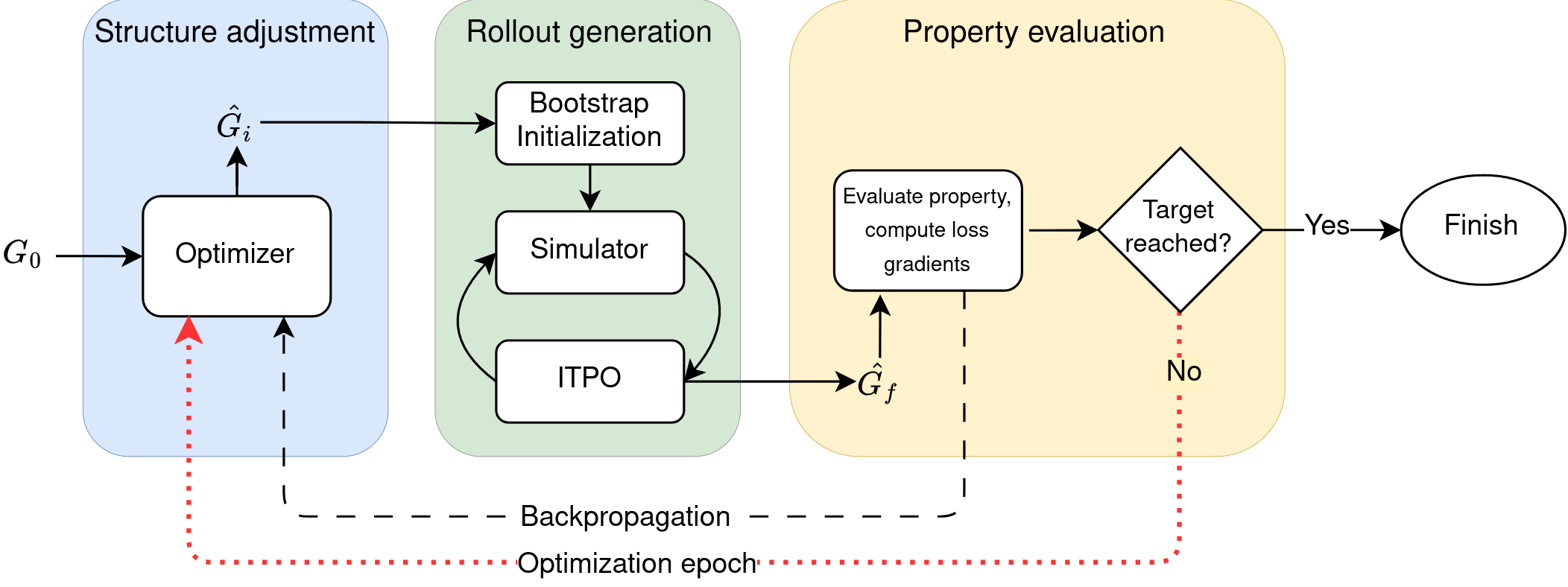}
  \caption{Schematic diagram of the proposed inverse design framework. In the structure adjustment stage (blue container), the initial graph $G_0$ is modified by the optimizer to yield the updated structure $\hat{G}_i$. During rollout generation (green container), historical context is generated via bootstrap initialization, and the GMDS coupled with inference-time physics-based optimization (ITPO) produces a rollout trajectory, shown with the final configuration $\hat{G}_f$. In the property evaluation stage (yellow container), the target dynamical property is evaluated to compute the loss gradients. The black dashed arrow represents gradient backpropagation from the objective function to the optimizer, while the red dotted line corresponds to the outer optimization loop.}
  \label{fig::opt_flowchart}
\end{figure}

The proposed inverse design pipeline embeds the GMDS within a gradient-based optimization framework. As illustrated in Fig. \ref{fig::opt_flowchart}, the optimization proceeds in a closed loop. First, specific modifications to a baseline system configuration are parameterized as learnable variables, such as node displacements or changes in harmonic bond stiffness. These modifications are applied to generate a candidate system, which is then provided to the GMDS to produce a rollout trajectory. The target macroscopic property, such as Poisson's ratio $\nu$ or the stress--strain response, is subsequently evaluated, and its exact gradient with respect to the learnable structural parameters is computed through backpropagation. The optimizer uses this gradient to update the structure in order to minimize the objective function. This cycle is repeated until either the target macroscopic behavior is achieved or the maximum number of optimization epochs is reached. A detailed technical description of the optimization procedure is provided in Section \ref{sec::comp::opt}.

The proposed framework is highly modular. Because the GMDS is the only system-specific component, while the remainder of the pipeline is system-agnostic, the approach can be readily generalized to the inverse design of diverse material and molecular systems. The objective function is similarly flexible, allowing multiple target properties and structural constraints to be combined as interchangeable components. For example, in addition to the primary optimization target, geometric penalties, such as constraints on minimum node distances or bond angles in elastic networks, can be incorporated directly into the loss function. Such terms help ensure the physical validity of the engineered structures and stabilize the optimization process without requiring changes to the overall architecture. 

\subsection{Graph Neural Network simulator}\label{sec::methods::gnn}

The core of the optimization pipeline is the GMDS model that was implemented following the established architecture \cite{Battaglia2018RelationalNetworks, Sanchez-Gonzalez2020, Pfaff2021, Salman2025EvaluatingNetworks, Shteingolts2026EnablingSimulators}. The objective of the GMDS is to auto-regressively generate trajectories by making consecutive single-step predictions based on a series of preceding system configurations.

The simulator takes the structure of a DEN at time $t$ as input, encoded into a graph $G^t$ with nodes $V$ (elastic network beads) and edges $E$ (harmonic bonds and other non-bonded interactions), along with their associated node features $\textbf{x}$ and edge features $\textbf{e}$. It then outputs predicted particle accelerations $\hat{\textbf{a}}^t$, which are used to calculate the next-step positions $\textbf{r}^{t+1}$ via semi-implicit forward Euler integration, where $\Delta t = 1$:
\begin{equation}
    \textbf{r}^{t+1} = \textbf{r}^t + \textbf{v}^{t} \Delta t +\hat{\textbf{a}}^t \Delta t ^ 2
\end{equation}
The number of prior system states included in the input graph defines the temporal history $h$. The simulator model used in this work utilizes a history of $h=3$, meaning that the node features $\textbf{x}$ encode the velocities of the current and two preceding timesteps ($\textbf{v}^t$, $\textbf{v}^{t-1}$, $\textbf{v}^{t-2}$). More details on the GNN architecture as well as specifics about the simulator model implementation are available in Section 3 of the SI.

\subsubsection{Dataset and Model training}\label{sec::methods::dataset_training}

The simulator training data consisted of MD simulations of uniaxial compression of DENs. The simulations were performed using the LAMMPS molecular dynamics package.\cite{Thompson2022} As part of the data preparation procedure, the compression trajectories were coarse-grained by a factor of 200, i.e., one GMDS model timestep equals 200 ground truth MD timesteps. This temporal coarse-graining allows the model to capture larger strains in fewer prediction steps, effectively reducing error accumulation without significantly diminishing per-step accuracy. Moreover, this enables a substantial advantage of the GMDS compared to existing differentiable MD engines, namely substantially reducing the complexity of gradient step tracking during optimization. To test the OOD generalization capabilities of the proposed inverse design framework, the GMDS was trained exclusively on compression trajectories of highly non-auxetic DENs with $\nu$ above a predefined threshold. Furthermore, the simulator training was additionally restricted to a strain $\epsilon$ of $0.0005$.

The GMDS was trained utilizing a multi-step training (MST) strategy. Instead of making isolated single-timestep predictions, for each training sample during the forward pass the model performs a short, continuous rollout. To maintain the training stability, the rollout length increases as a function of training epochs, starting at a single step and scaling up to 10 steps in later epochs. By supervising across this short rollout, the simulator model is exposed to its own state drift, effectively encouraging it to learn ``self-correction'', which significantly reduces error accumulation (see Section 3.2 of the SI). A detailed description of the GMDS training protocol is provided elsewhere.\cite{Shteingolts2026EnablingSimulators}

The GMDS model was trained on three distinct datasets. Each of the three datasets was composed of DENs compression trajectories, spanning a wide range of Poisson's ratios $\nu \in [\approx -0.6, 0.6]$. All datasets were created with a two-step process. First, a set of unique, non-auxetic DENs was generated. These networks were then optimized using one of three optimization algorithms. The first dataset was created using the gradient-descent-based global node displacement strategy.\cite{Shen2024} For the second dataset, we employed a variation of the harmonic bond stiffness tuning algorithm,\cite{Reid2018} which relies on iteratively pruning individual harmonic bonds to minimize the ratio between network's shear and bulk modulus $\frac{G}{B}$. Lastly, we also generated a dataset of DENs using a random perturbation strategy, which involves repeatedly applying a random displacement to a certain percentage of nodes, as well as randomly pruning some of the network bonds, until it reaches a desired Poisson's ratio value. The DENs generated using the random perturbation strategy were additionally augmented with non-bonded Lennard--Jones interactions, which we found to lead to more complex compression dynamics. Detailed description of the datasets generation is available in Section 2 of the SI.

\subsection{Structure-only initialization}

As described in the Introduction, one of the main challenges in employing dynamic ML-based simulators for inverse design is their reliance on dynamical state variables, such as velocities and accelerations, at the current timestep $t$ as well as at preceding timesteps. In practice, because the model trajectory is coarse-grained in time, these quantities are typically computed from preceding reference frames. Consequently, the input to a typical ML-based simulator requires historical context (see Section \ref{sec::methods::gnn}). Previously, we presented two potential solutions to this structure-only initialization (SOI) problem.\cite{Shteingolts2026EnablingSimulators}

In this work, we employ a custom-built, differentiable MD engine implemented using PyTorch \cite{Paszke2019, Ansel2024PyTorchCompilation} and PyTorch Geometric \cite{Fey2019, Fey2025PyGGraphs} to generate a short ``bootstrap'' trajectory from a single static input configuration. Specifically tailored for uniaxial compression of 2D DENs in the NPT ensemble, this MD engine generates a transient trajectory of length $l = h \cdot d_{cg}$, where $h$ denotes the required model history and $d_{cg}$ is the coarse-grained timestep of the GMDS. By extracting the required sequence of configurations from this short simulation, we establish the initial dynamical context needed to initialize the simulator.

\subsection{Inference-time physics-based optimization}

Rollout stability and OOD generalization of the GMDS model are significantly enhanced by employing Inference-Time Physics-based Optimization (ITPO).\cite{Shteingolts2026EnablingSimulators} ITPO treats the GMDS's predicted accelerations as a high-quality initial approximation, which is subsequently refined by enforcing predefined physical constraints. This continuous trajectory correction noticeably improves the simulator's performance in extrapolative regimes. In this work, the ITPO refinement is applied at every rollout step; however, it can in principle be applied every $N$ steps to reduce the computational cost. While a detailed description of the ITPO technique is available elsewhere,\cite{Shteingolts2026EnablingSimulators} we provide a brief overview here.

The GMDS predicted accelerations $\hat{\textbf{a}}$ are refined via gradient-based optimization using the following composite loss function: 
\begin{equation}
    \mathcal{L}_{\text{total}} = \mathcal{L}_{\text{anchor}} + \alpha \mathcal{L}_{\text{physics}}
\end{equation}
where $\alpha$ is an empirical weighting coefficient. The anchor term $\mathcal{L}_{\text{anchor}}$ penalizes deviations from the initial prediction to preserve the learned compression dynamics. In contrast, $\mathcal{L}_{\text{physics}}$ acts as a corrective penalty, steering the system toward a state that is consistent with the predefined physical constraints. For the uniaxial compression of DENs, $\mathcal{L}_{\text{physics}}$ combines three domain-specific constraints: a barostat pressure penalty, the system potential energy, and the mean-squared per-particle force, briefly described below.

\textit{Barostat Term:} Given a constant engineering strain applied along the $x$ axis, a differentiable GMDS-coupled barostat regulates the instantaneous transverse internal pressure $P_y$. This pressure comprises kinetic and virial contributions:
\begin{equation}
P_y = \frac{k_B NT}{A} + \frac{1}{A} \sum_{i < j} f_{ij, y} r_{ij, y} \label{eq::bar1}
\end{equation}
where $A$ is the simulation box area, $f_{ij, y}$ is the $y$-component of the harmonic bond force between nodes $i$ and $j$, and $r_{ij, y}$ is their relative distance. The deviation of $P_y$ from the target external pressure, $P_t$, generates a thermodynamic driving force:
\begin{equation}
    F_{\text{total}} = (P_y - P_t) L_x - \gamma v^t_y \label{eq::bar2}
\end{equation}
with $\gamma$ serving as a frictional damping coefficient. The resulting box acceleration, $a^t_y = F_{\text{total}}/W_y$ (where $W_y$ is the system piston mass), continuously updates the transverse box dimension, $L_y$:
\begin{eqnarray}
    v^{t}_y = v^{t-1}_y  + a^t_y \Delta t \\
    L^{t+1}_y = L_y^t + v^{t}_y \Delta t \label{eq::bar3}
\end{eqnarray}

\textit{Potential Energy Term:} Because non-physical trajectories tend to exhibit an artificial inflation of system energy, we minimize the total potential energy, formulated as the sum of all harmonic bond energies:
\begin{equation}
    \mathcal{L}_U = \sum_{i<j} E(\textbf{r}_{ij})
\end{equation}

\textit{Per-Particle Net Force Term:} Similarly, compounding rollout errors often manifest as large per-particle forces. To counteract this, we penalize the mean-squared net force along the $y$ axis:
\begin{equation}
    \mathcal{L}_{\text{force}} = \frac{1}{N} \sum_{i=1}^N (F_{i, y})^2
\end{equation}
where the net force $\mathbf{F}_i$ acting on node $i$ is derived from its neighboring bonds:
\begin{equation}
    \mathbf{F}_i = \sum_{j \in \mathcal{N}(i)} -k_{ij} (l - l_0) \frac{\mathbf{r}_{ij}}{l}
\end{equation}
Here, $k_{ij}$ is the harmonic bond stiffness, while $l$ and $l_0$ represent the current and bond rest lengths, respectively.

\section{Results}

To evaluate the proposed framework's capacity for OOD inverse design and optimization, we examined three distinct scenarios: \textit{i}) designing DENs with Poisson's ratios substantially lower than those represented in the training dataset, \textit{ii}) optimizing DENs for a target stress--strain response and \textit{iii}) introducing local inhomogeneities or ``defects'', in the form of stress concentrations and localized variations in Poisson's ratio. 

\subsection{Inverse Design of Auxetic Networks from High Poisson's Ratio Configurations }

Inverse design of auxetic DENs provides a useful case study for evaluating the OOD capabilities of the proposed dynamical optimization framework. As shown previously,\cite{Salman2025EvaluatingNetworks} the compression trajectories of auxetic DENs exhibit substantially greater complexity than those of non-auxetic networks. Restricting the GMDS training data exclusively to non-auxetic, high-$\nu$ compression trajectories therefore provides a meaningful OOD benchmark. Specifically, the simulator used in this study was trained only on data from networks with $\nu \geq 0.1$. Despite this restriction, the framework consistently and reliably engineers auxetic DENs with negative Poisson's ratios as low as $\nu=-0.3$, starting from baseline configurations with $\nu \geq 0.35$.

\subsubsection{Node displacement optimization}

\begin{figure}[ht]
  \centering
  \includegraphics{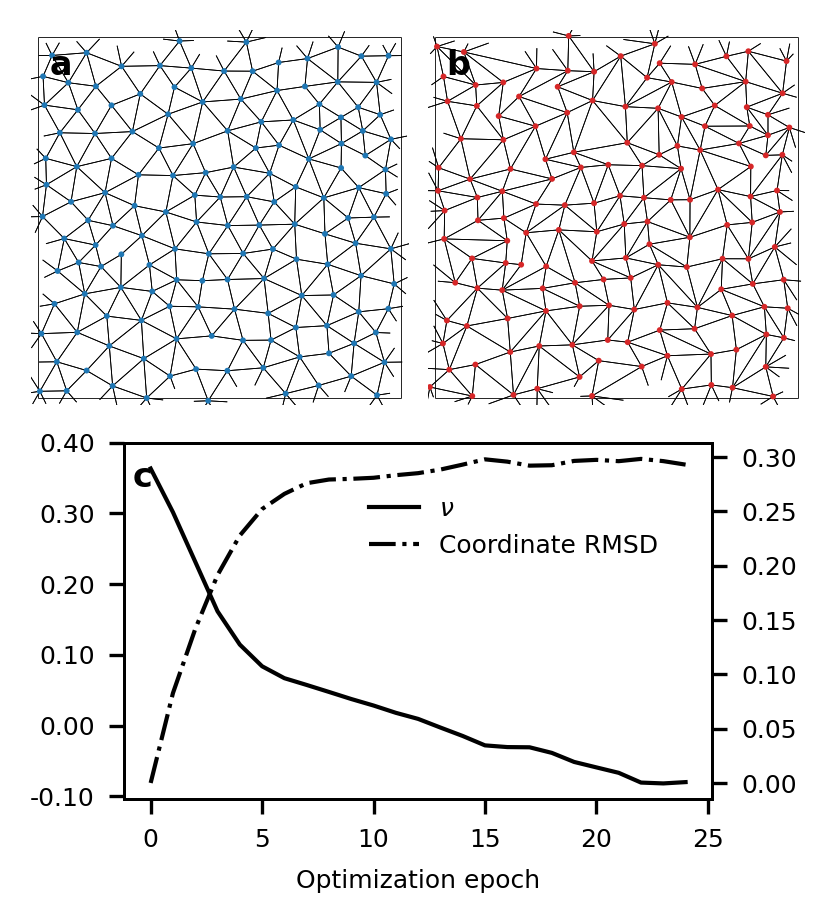}
  \caption{A disordered elastic network before (a) and after (b) optimization of its Poisson's ratio, together with the optimization trajectory (c), showing $\nu$ and the root-mean-square deviation (RMSD) of the node positions as functions of the optimization epoch.}
  \label{fig::disp_opt}
\end{figure}

As demonstrated elsewhere \cite{Shen2024}, modifying node positions enables effective engineering of DENs with targeted Poisson's ratios. The optimization module within our framework employs a similar strategy. Specifically, the Poisson's ratio $\nu$ of a network is minimized by first modifying node positions and then recalculating the harmonic bond stiffnesses $k$ so that the resulting configuration remains in mechanical equilibrium. Figure \ref{fig::disp_opt} shows an example of an optimized network topology. Using the dynamical optimization framework, we managed to consistently engineer DENs with Poisson's ratios as low as $\nu = -0.10$ starting from networks with $\nu \geq 0.35$. 

\begin{figure}[ht]
  \centering
  \includegraphics{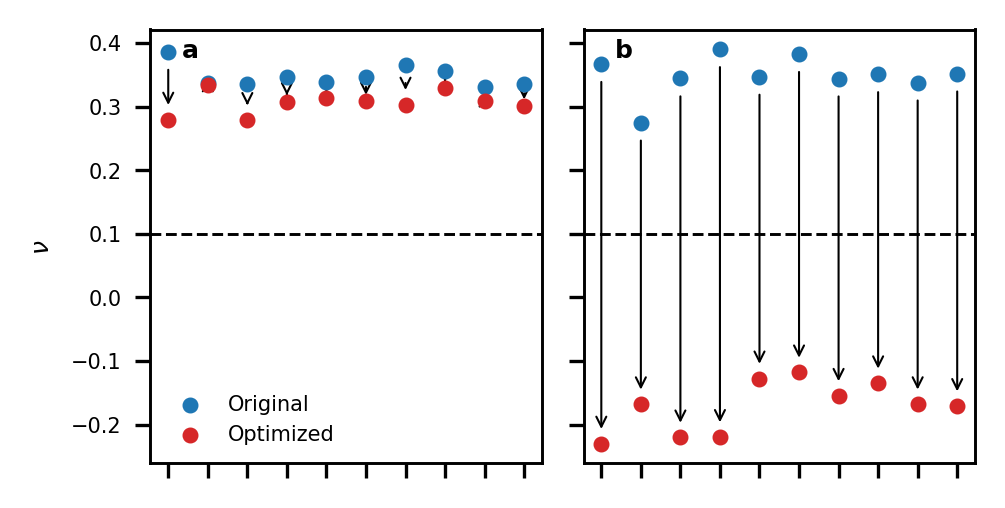}
  \caption{Randomly selected examples of Poisson's ratio optimization of DENs via node displacement, using the proposed inverse design framework with the static $\nu$-predictor model (a) and the GMDS (b). Blue and red dots correspond to the values of $\nu$ before and after optimization, respectively. The training-data cutoff in $\nu$ for both models is indicated by the black dashed line.}
  \label{fig::dyn_static_opt_comparison}
\end{figure}

As noted in the Introduction, static structure-based surrogate models are known to exhibit limited generalization beyond their training distribution. To quantify this limitation and establish a baseline for comparison, we trained a comparable Encoder--Processor--Decoder model to predict the Poisson's ratio $\nu$ directly from the static input graph $G$, which encodes the DEN node coordinates $\mathbf{r}^0$ and edges $\mathbf{e}^0$ prior to compression. The model exhibited poor OOD generalization (Fig. S1) and, when incorporated into the gradient-based inverse design pipeline (Fig. \ref{fig::opt_flowchart}), failed to generate valid structures with low Poisson's ratio in the extrapolative regime (Fig. \ref{fig::dyn_static_opt_comparison}).

Notably, the proposed dynamical optimization framework can identify optimized solutions while simultaneously satisfying structural constraints. For instance, when using the node-displacement strategy, maintaining a physically valid network structure requires geometric constraints that penalize undesirable configurations. Following previous work,\cite{Shen2024} we imposed two such constraints: the minimum length of each bond was constrained to remain at least 30\% of its original value, $l_\text{min} = 0.3\;l_0$, and the angle between any two adjacent harmonic bonds was constrained to remain at least $\theta_\text{min} = 20 \degree$.

To further test both the OOD limits of the GMDS and the flexibility of the inverse design framework, we introduced additional structural constraints, including an upper bound on the RMSD to limit the overall displacement of the network nodes. As illustrated by the optimization trajectory in Figure \ref{fig::disp_opt}c, the RMSD rapidly approaches its prescribed upper bound at approximately $0.3$, while the Poisson's ratio $\nu$ continues to decrease before eventually reaching a plateau toward the end of the optimization.

\begin{figure}[ht]
  \centering
  \includegraphics{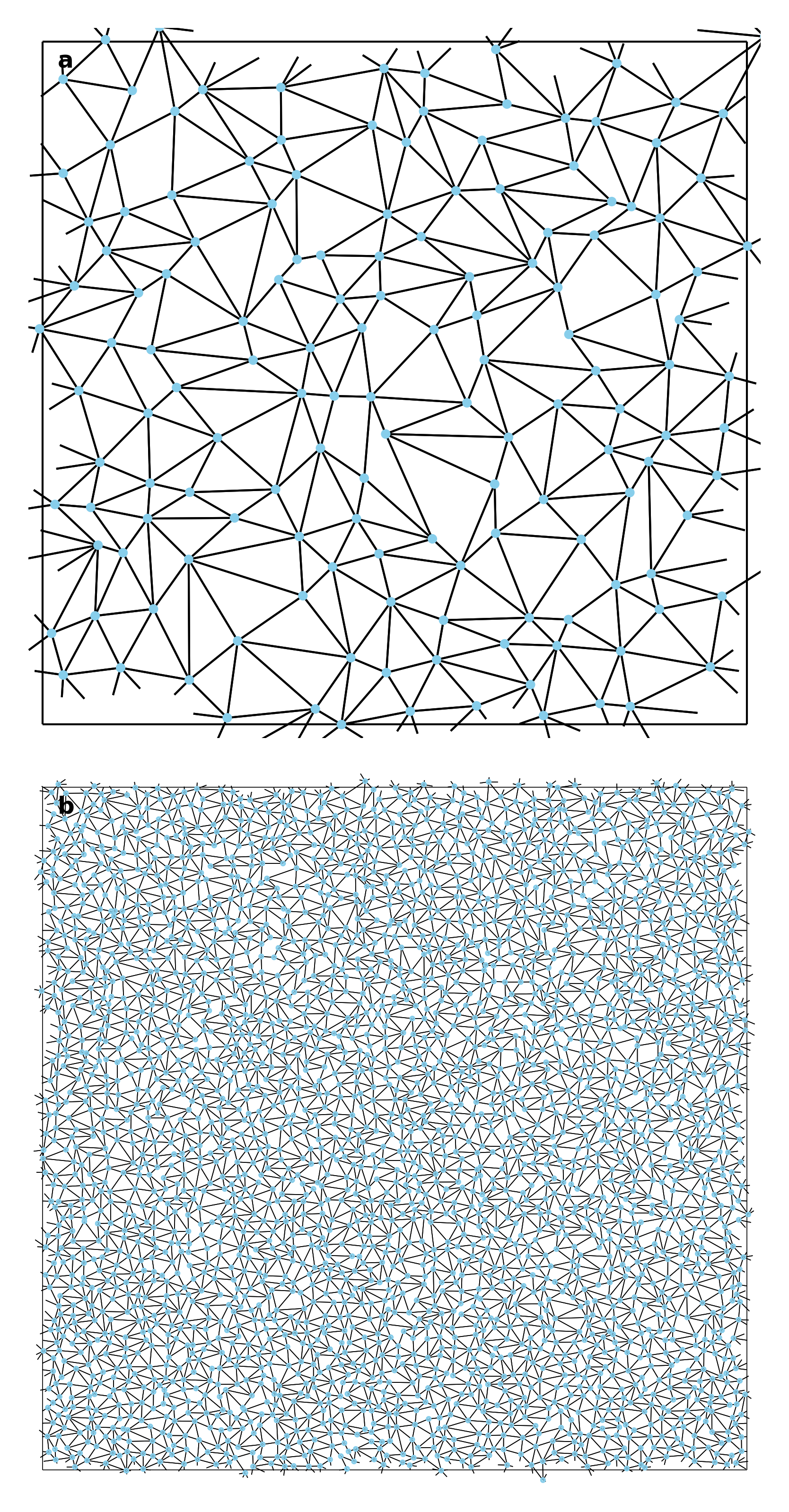}
  \caption{Examples of auxetic DENs with approximately 150 (a) and 3000 nodes (b), obtained using the proposed inverse design framework. The GMDS training data consisted of networks with 150--200 nodes.}
  \label{fig::size_generalization}
\end{figure}

A further demonstration of the proposed framework's OOD capabilities is its ability to generalize across varying system sizes. Despite the fact that the GMDS was trained exclusively on DENs with fewer than 200 nodes, the framework reliably engineered auxetic networks with up to 5000 nodes without modifying the optimization procedure (Fig. \ref{fig::size_generalization}). Because the computational cost of optimizing larger DENs is dominated by the GMDS rollout, the proposed approach scales significantly better than conventional greedy algorithms, making generalization to larger systems advantageous for large-scale inverse design.

\subsubsection{Bond stiffness tuning}

We were also interested in evaluating the framework's ability to perform OOD design through bond stiffness $k$ modifications. Previous studies \cite{Goodrich2015, Hexner2018, Hexner2018a, Reid2018, Reid2019, Hexner2020} have shown that selective bond pruning using greedy algorithms can reduce the Poisson's ratio of DENs. Motivated by these findings, we investigated whether the proposed inverse design framework could achieve similar reductions through continuous adjustment of $k$. As in the previous experiments, the GMDS was trained exclusively on data from networks characterized by $\nu \geq 0.1$. Again, the framework proved highly effective at OOD optimization. 

\begin{figure}[ht]
  \centering
  \includegraphics{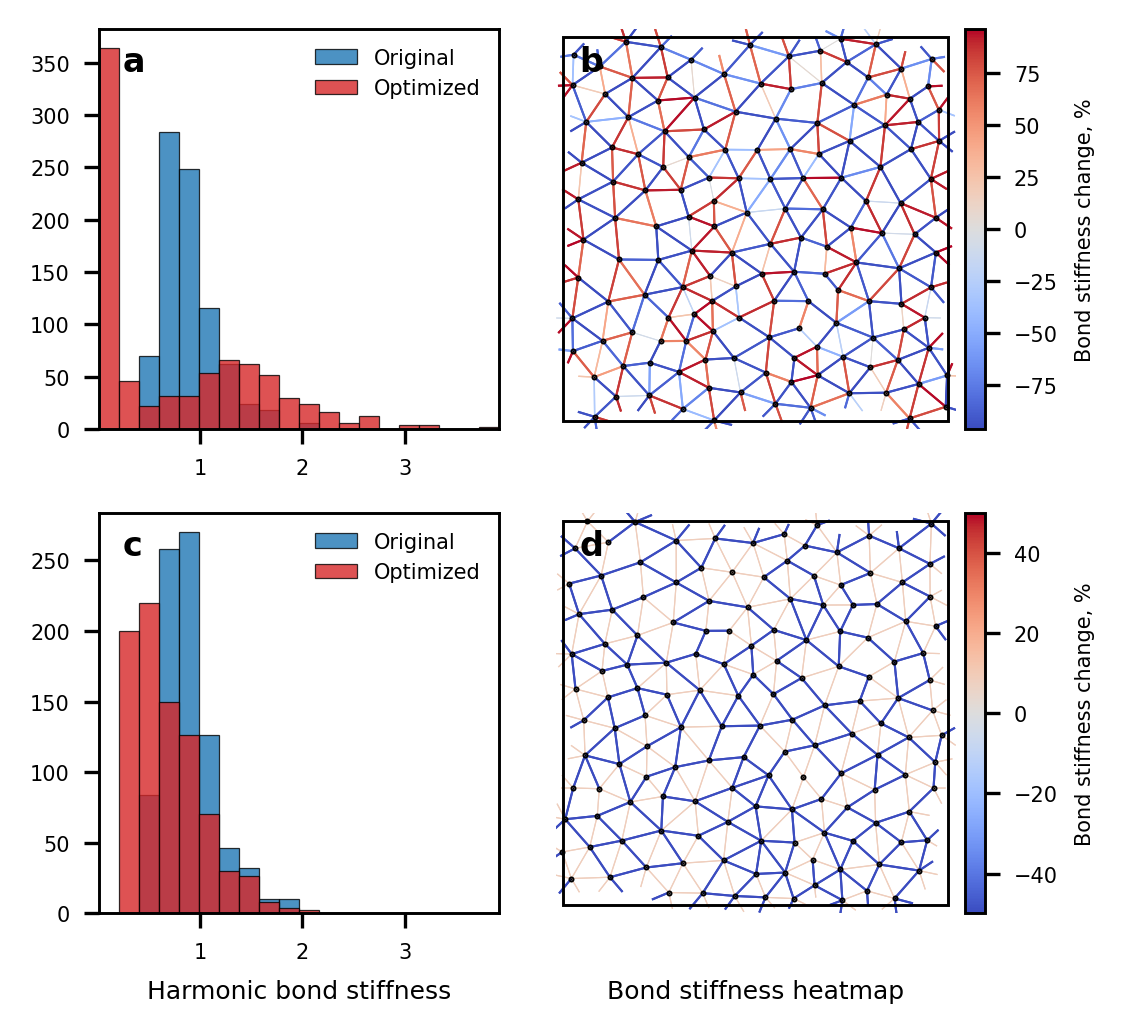}
  \caption{Distributions of harmonic bond stiffnesses $k$ before (blue) and after (red) optimization using bond stiffness tuning (a, b) and the combined strategy (c, d). Weakened and strengthened bonds are shown with blue and red, respectively.}
  \label{fig::stiff_opt}
\end{figure}

In contrast to conventional greedy algorithms, which rely on a binary pruning decision, i.e., setting the stiffness to $k=0$ or retaining its original value, optimizing a continuous stiffness parameter enables a more granular exploration of the design space. As illustrated in Fig.~\ref{fig::stiff_opt}, the optimized bond stiffness distribution differs substantially from the original one. Following optimization, the original distribution $k=1/l_0$ was replaced by two distinct peaks: approximately $50\%$ of the harmonic bonds had $k\approx 0$, while a second local maximum appeared near $k\approx 1.15$. 

Compared to the node displacement strategy, optimizing individual bond stiffnesses is procedurally more straightforward, as it eliminates the need to impose structural constraints. Instead, the optimization is subject only to upper and lower bounds on the bond stiffness. In this case, the stiffness $k_{ij}$ of each bond was constrained to deviate by no more than $90\%$ from its original value.

We also examined a joint optimization strategy combining node displacement and bond stiffness tuning. This strategy yielded lower values of $\nu$ for a given elastic network. The Poisson's ratio reached values as low as $-0.35$, compared with the minimum of approximately $-0.1$ typically obtained when either strategy was applied independently. In addition to yielding more strongly auxetic DENs, the joint optimization strategy provided a significant computational speedup, substantially reducing the number of epochs required to achieve a comparable negative Poisson's ratio.

As shown by the bond stiffness distribution in Fig.~\ref{fig::stiff_opt}c and d, in the absence of additional structural constraints, within the joint optimization strategy the initial stiffness $k$ of approximately half of the harmonic bonds tended to be uniformly reduced by $40\%$, while for the remaining bonds $k$ was left largely unchanged.

\subsection{Modifying the stress--strain curve}

\begin{figure}[ht]
  \centering
  \includegraphics[width=3.33in]{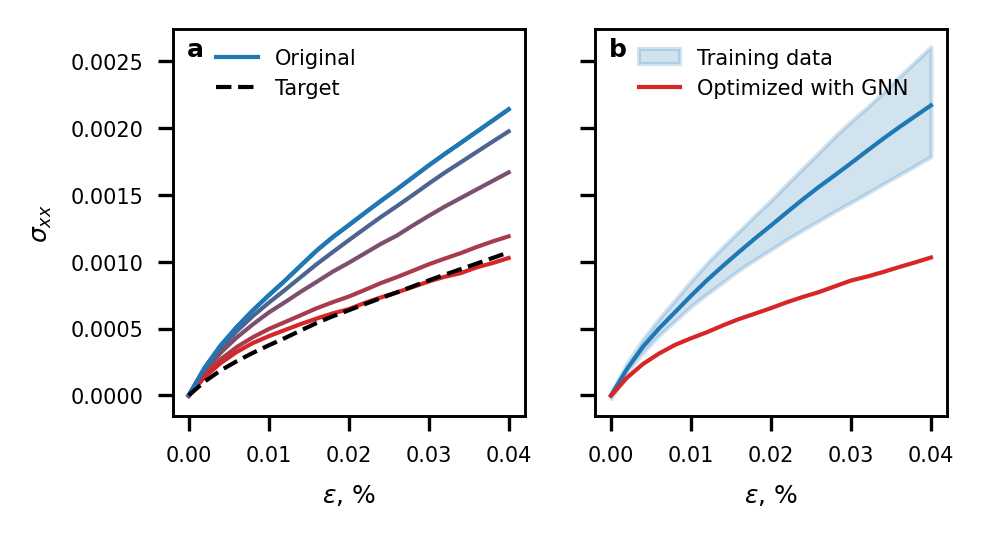}
  \caption{Optimization of the stress--strain response of DENs with additional non-bonded Lennard--Jones interactions: (a) evolution of the stress--strain curve during optimization; (b) comparison of the optimized response with those of DENs in the training dataset. In (a), successive curves are shown at intervals of 5 optimization epochs.}
  \label{fig::stress}
\end{figure}

An alternative optimization target is the direct modification of the material's macroscopic stress--strain response. In the small-strain regime ($\epsilon \leq 0.01$), unmodified baseline DENs exhibit an almost perfectly linear stress--strain relationship. By contrast, the introduction of non-bonded Lennard--Jones interactions produces a more complex, nonlinear response (Fig.~\ref{fig::stress}a). Here, we investigated whether the proposed framework could engineer targeted mechanical responses in both cases. The inverse design framework successfully reshaped both the linear and nonlinear baseline stress--strain curves toward predefined targets. Continuous bond stiffness tuning produced the most robust results for this task, enabling shifts in the stress--strain response of up to $50\%$ relative to the unoptimized baseline. Notably, as shown in Fig.~\ref{fig::stress}b, the optimized stress--strain curve lies well outside the range of responses represented in the training dataset.

\subsection{Engineering local defects}

Beyond optimizing global mechanical properties, the proposed framework can also engineer localized defects within an existing elastic network. Specifically, we demonstrated that the framework could create local pockets whose dynamic responses differed substantially from those of the surrounding structure while leaving the global mechanical properties largely intact. We considered two examples of engineered local pockets: \textit{i}) a region exhibiting a negative local Poisson's ratio $\nu$ and \textit{ii}) a region exhibiting a localized stress concentration, characterized by $\sigma_{xx}^{\text{local}}$. In both cases, the pocket was defined by a radial cutoff around a central node and typically contained $10\%$ to $20\%$ of the network's nodes. By applying the optimization strategies discussed above exclusively to the nodes and/or bonds within the designated pocket, we introduced the targeted local behavior while largely preserving the network's global mechanical response.

\begin{figure}[ht]
  \centering
  \includegraphics[width=3.33in]{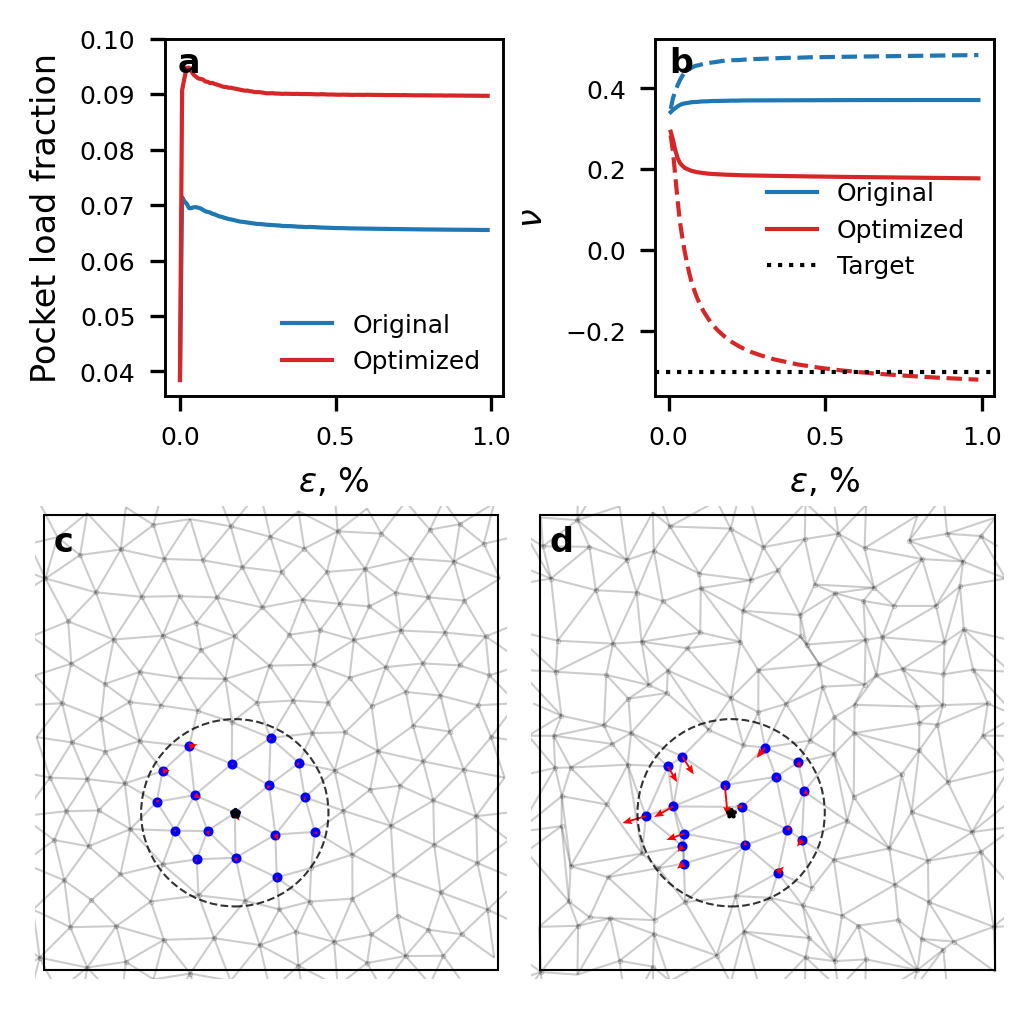}
  \caption{Examples of engineered local defects. (a) Increasing the load fraction $\gamma$ within a pocket. (b) Engineering a locally auxetic pocket; solid and dashed lines show the global and local Poisson's ratios $\nu$ as functions of strain, respectively. (c, d) Non-affine node displacement vectors before (c) and after (d) the introduction of an auxetic pocket. Results for the original and optimized structures are shown in blue and red, respectively.}
  \label{fig::pockets}
\end{figure}

Figure~\ref{fig::pockets}a shows the results of introducing a high-stress pocket while keeping the global stress $\sigma_{xx}$ approximately unchanged. After optimization, the stress within the targeted pocket increased more rapidly with strain, while the macroscopic network stress remained largely unchanged. To quantify this localization, we computed the load fraction $\gamma$, defined as the fraction of the total network stress carried within the pocket. Optimization increased $\gamma$ by nearly a factor of two to approximately $0.095$, indicating a substantial redistribution of load toward the targeted region.

The embedding of an auxetic pocket is illustrated in Fig.~\ref{fig::pockets}b. A typical unoptimized DEN exhibits a positive global Poisson's ratio $\nu \approx 0.35$, which remains approximately constant with strain $\epsilon$. Before optimization, both the global response $\nu_{\text{global}}$ and the local pocket response $\nu_{\text{pocket}}$ exhibited this behavior. After targeted optimization, $\nu_{\text{global}}$ decreased to approximately $0.2$, whereas $\nu_{\text{pocket}}$ reached approximately $-0.3$. This contrast demonstrates that the optimization framework can strongly modify the mechanical response of a localized region while inducing comparatively modest changes in the global behavior of the network. The resulting changes in node dynamics are illustrated in Fig. \ref{fig::pockets}c and d. Consistent with the behavior represented in the GMDS training set, node motion in the original DEN (Fig.~\ref{fig::pockets}c) was predominantly affine, with minimal non-affine displacements. After the auxetic pocket was introduced, the local node dynamics became substantially more complex (Fig.~\ref{fig::pockets}d), further illustrating the framework's ability to generate dynamical patterns absent from the training data.

Notably, localized defects of this type are entirely absent from the GMDS training data. The ability to generate such responses suggests that the simulator captures node-level dynamics in sufficient detail to resolve how localized structural modifications influence the overall compression trajectory. More broadly, these results provide further evidence that the framework supports OOD optimization, enabling the generation of network structures and mechanical responses not represented in the training data.

\section{Conclusion}

In this work, we introduced an inverse design framework for dynamics-based optimization. At its core, the framework leverages an autoregressive graph neural network molecular dynamics simulator (GMDS), a fully differentiable surrogate model that accurately predicts dynamic trajectories while exhibiting significant out-of-distribution (OOD) generalization. Through the simulated compression of disordered elastic networks (DENs), we demonstrated that this methodology can both engineer specific macroscopic, network-wide properties and embed localized regions with targeted dynamical behavior that are entirely absent from the original training data.

At the global scale, the framework produced highly auxetic networks with Poisson's ratios as low as $\nu=-0.3$, despite the GMDS having been trained exclusively on networks with $\nu\in[0.1,0.4]$. This corresponds to a shift of $\Delta\nu\approx0.7$ relative to a typical unoptimized DEN with $\nu\approx0.4$. The framework also enabled targeted modification of macroscopic stress--strain curves, yielding responses well beyond the training distribution. At the local scale, it embedded regions with distinct dynamical behaviors within otherwise standard non-auxetic networks, including isolated auxetic pockets and stress-concentrating regions that doubled the local load fraction $\gamma$ relative to the surrounding network.

These results were enabled primarily by two key advances: solving the structure-only initialization (SOI) problem and applying the Inference-Time Physics-Based Optimization (ITPO) technique. The SOI problem is addressed by generating a short transient trajectory, which provides the initial historical context required to initialize GMDS rollout. This, in turn, allows the gradient-based inverse design loop to start from a newly proposed static system configuration. ITPO further improves the OOD capabilities of the GMDS by refining the model's predictions so that the resulting configuration remains physically valid. Additionally, ITPO substantially improves the stability of GMDS rollouts. Together, these two techniques enable the GMDS to generate accurate rollout trajectories even when used to design systems with dynamical behavior significantly different from that encountered in its training dataset.

Compared with general-purpose differentiable MD frameworks \cite{Schoenholz2021, Doerr2021TorchMD:Simulations, Ple2024FeNNol:Potentials, Christiansen2025FastSimulations, Cohen2025TorchSim:PyTorch}, which enable targeted engineering of the underlying interactions but frequently incur high computational costs and suffer from exploding or high-variance gradients when differentiating through many timesteps \cite{Ingraham2019LearningSimulator, Thaler2021LearningReweighting, Metz2022GradientsNeed, Sipka2023DifferentiableEvents, Greener2024DifferentiableProteins}, the proposed approach offers a complementary route in which temporally coarse-grained dynamics can reduce both computational cost and the depth of the differentiable computational graph. Thus, by requiring fewer sequential gradient operations and filtering fast dynamical modes, this formulation can improve the conditioning and stability of long-horizon gradient-based optimization.

Ultimately, this work establishes a foundation for the use of fully differentiable GNN-based MD simulators in the inverse design of complex systems. By demonstrating that such models can guide the optimization of dynamical properties well beyond their training distributions, our results highlight the potential of learned simulators as efficient tools for targeted structural design and materials discovery. Future work will aim to extend this framework to a broader range of dynamical optimization problems and to increasingly complex molecular and materials systems.

\section{Computational details}

\subsection{Autoregressive rollout and ITPO refinement}

The generation of the coarse-grained compression trajectory is handled via an autoregressive loop that integrates the GMDS and the ITPO. The process is bootstrapped by generating a transient sequence of $h+1$ configurations using a custom differentiable MD engine. Implementation details are available in Section 5 of the SI. Once the initial dynamical context is established, the GMDS predicts the baseline particle accelerations $\hat{\mathbf{a}}_\text{nn}$ for each subsequent timestep. The ITPO technique is used to refine the simulator's prediction. To prevent excessive memory consumption during training and optimization, the forward refinement pass is completely detached from the global computational graph. The ITPO loop utilizes the Adam optimizer to iteratively minimize the composite physical loss, $\mathcal{L}_{\text{total}}$, yielding the refined particle accelerations $\mathbf{a}_\text{nn}^*$, which are then used to compute the subsequent system configuration $\mathbf{r}^{t+1}$. For details on ITPO implementation see Section 6 of the SI.

\subsection{Detailed description of the optimization algorithm}\label{sec::comp::opt}

\subsubsection{Structural parameterization}

The optimization seeks to minimize an objective function by updating a set of learnable structural parameters. Depending on the chosen strategy, the structural modifications are parameterized as either node displacements $\Delta \mathbf{r}$ or bond stiffness adjustments $\Delta \mathbf{k}$.

For global node displacement, a learnable tensor of the same shape as the initial node positions is initialized. At each optimization epoch, the updated node coordinates are computed as $\mathbf{r}_i = \mathbf{r}_i^0 + \Delta \mathbf{r}_i$. To prevent global translation of the network during optimization, the center of mass of the displacement tensor is subtracted at every step.

For bond stiffness tuning, enforcing physical bounds on the optimized stiffness values is important to prevent unrealistically high values of $k_{ij}$. First, a learnable adjustment tensor $\mathbf{w}$ corresponding to the network edges is initialized. Because the GMDS operates on bidirectional graphs, to guarantee symmetry $\mathbf{w}_{ij} = \mathbf{w}_{ji}$ the tensor is averaged with its transpose. A bounded stiffness multiplier $m_{ij}$ is then constructed using a hyperbolic tangent activation:
\begin{equation}
    m_{ij} = \exp \left( S \cdot \tanh(\mathbf{w}_{ij}) \right)
\end{equation}
where $S$ is a scaling factor defined by the logarithmic upper and lower bounds of the allowed multiplier. e.g., between $10^{-6}$ and $5.0$. The updated bond stiffness is then calculated as 
\begin{equation}
    k_{ij} = k_{ij}^0 \cdot m_{ij}
\end{equation}
ensuring that $k_{ij}$ remains within the predefined physical limits.

\subsubsection{Forward pass and loss formulation}

During the forward pass, the input graph is passed to the GMDS to generate the rollout trajectory. We found that in most cases using 20-step long rollouts produces satisfactory results. Once the rollout trajectory is generated, the target dynamical property is evaluated and the composite loss function $\mathcal{L}_{\text{total}}$ is computed. This loss acts as the optimization objective and is defined as a sum of the target property term $\mathcal{L}_\text{target}$ and the optional geometric constraints term $\mathcal{L}_\text{gc}$:
\begin{equation}
    \mathcal{L}_\text{total} = \mathcal{L}_\text{target} + \mathcal{L}_\text{gc}
\end{equation}
The target property term $\mathcal{L}_{\text{target}}$ penalizes the deviation of the engineered property from the target value. For the auxetic optimization this loss term is computed as the Mean Squared Error (MSE) between the current and the target Poisson's ratio $\nu$:
\begin{equation}
    \mathcal{L}_\text{target} = \left( \nu - \nu_\text{target} \right)^2    
\end{equation}
Poisson's ratio $\nu$ is defined as a negative ratio of transverse strain to the axial strain:
\begin{equation}
    \nu = -\frac{d\epsilon_y}{d\epsilon_x}
\end{equation}
For a given GMDS rollout, transverse $\epsilon_y$ and axial $\epsilon_x$ strain are computed from the periodic box dimensions:
\begin{equation}
    \epsilon_{\eta} = (L^0_{\eta} - L_{\eta}) / L_{\eta}
\end{equation}
where $\eta \in [x, y]$ and $L_{\eta}$ corresponds to the periodic box dimension.

For the stress--strain curve engineering, a shape-aware target loss is employed that explicitly constrains the first- and second-order derivatives of the curve. Let $\mathbf{S}$ and $\mathbf{S}^*$ be the predicted and target stress--strain curves, respectively. The target loss is formulated as a linear combination of positional $\mathcal{L}_\text{pos}$, slope $\mathcal{L}_\text{slope}$, and curvature $\mathcal{L}_\text{curve}$ penalties:
\begin{equation}
    \mathcal{L}_\text{target} = \mathcal{L}_\text{pos} + \alpha \mathcal{L}_\text{slope} + \beta \mathcal{L}_\text{curve}   
\end{equation}
where $\alpha$ and $\beta$ are empirical weighting coefficients. The slope term $\mathcal{L}_\text{slope}$ and the curvature term $\mathcal{L}_\text{curve}$ are the MSE of the first-order and second-order derivatives, respectively, computed using the finite differences.

A localized pocket is defined by a central target node $c$ and a set of neighboring nodes $\mathcal{N}_c$ located within a specified radius. To evaluate the local Poisson's ratio $\nu_\text{local}$, the initial node positions $\mathbf{r}^0$ and the final node positions at the end of the rollout trajectory $\mathbf{r}^t$ are used. The relative distances along the $x$ and $y$ axes between the central node and its neighbors are computed for both states:
\begin{equation}
    \Delta x_i = | x_i - x_c |, \\
    \Delta y_i = | y_i - y_c |
\end{equation}
where $i \in \mathcal{N}_c$. The macroscopic pocket strains $\epsilon_x^{\text{local}}$ and $\epsilon_y^{\text{local}}$ are then calculated by comparing the average final distances to the average initial distances:
\begin{equation}
    \epsilon_x^\text{local} = \frac{\langle\Delta x_i^{t}\rangle}{\langle\Delta x_i^0\rangle} - 1 \\
    \epsilon_y^\text{local} = \frac{\langle\Delta y_i^{t}\rangle}{\langle\Delta y_i^0\rangle} - 1
\end{equation}
Given a target local Poisson's ratio $\nu_{\text{target}}^{\text{local}}$, the desired local $y$ strain $\epsilon_{y, \text{target}}^{\text{local}}$ is defined as
\begin{equation}
    \epsilon_{y, \text{target}}^{\text{local}} = -\nu_{\text{target}}^{\text{local}} \cdot \epsilon_x^{\text{local}}   
\end{equation}
The defect target loss term $\mathcal{L}_{\text{defect}}$ is then computed as the MSE between the observed and the calculated target $y$ strains:
\begin{equation}
    \mathcal{L}_{\text{defect}} = \lambda \left( \epsilon_y^{\text{local}} - \epsilon_{y, \text{target}}^{\text{local}} \right)^2
\end{equation}
We found that expressing the local defect target loss in terms of local $y$ strain stabilized the optimization significantly.

\subsubsection*{Structural constraints}

To maintain the physical validity of the network structure, particularly during node displacement optimization, the geometric constraints loss $\mathcal{L}_\text{gc}$ is used, which consists of two separate terms $\mathcal{L}_\text{dist}$ and $\mathcal{L}_\text{angles}$. The distance penalty $\mathcal{L}_{\text{dist}}$ prevents nodes from overlapping by penalizing bond lengths $d_{ij}$ that fall below a specified minimum threshold $d_{\min}$ using the ReLU function:
\begin{equation}
    \mathcal{L}_{\text{dist}} = \sum_{E} \left[ \text{ReLU}(d_{\min} - d_{ij}) \right]^2    
\end{equation}
Similarly, the angle loss term $\mathcal{L}_{\text{angle}}$ prevents intersecting bonds. It penalizes angles $\theta_{ijk}$ that fall below a minimum threshold $\theta_{\min}$ using a Softplus function with $\beta = 3.0$:
\begin{equation}
    \mathcal{L}_\text{angle} = \sum_{V} \text{Softplus}(\theta_{\min} - \theta_{ijk}, \beta)
\end{equation}

\subsubsection*{Optimization and backpropagation}

The structural parameters are updated using the Adam optimizer \cite{Kingma2017Adam:Optimization} with the learning rate set to 0.05 and no weight decay. To ensure stability, gradient clipping is applied by capping the maximum $L_2$ norm of the gradients at 1.0 before the optimizer step is executed. 

\subsection{Load fraction}

The total load $\mathcal{W}$ is defined as:

\begin{equation}
    \mathcal{W} = \sum |w_{ij}|
\end{equation}
where $w_{ij}$ is the local virial stress contribution corresponding to the harmonic bond between particles $i$ and $j$:
\begin{equation}
    w_{ij} = \mathbf{r}_{ij} \cdot \mathbf{F}_{ij} = r_xF_x + r_yF_y
\end{equation}
The load fraction carried by an engineered pocket is thus defined as:
\begin{equation}
    \gamma = \frac{\mathcal{W}_\text{pocket}}{\mathcal{W}_\text{total}}
\end{equation}

\section*{Acknowledgements}

The authors acknowledge support from the Israel Science Foundation (ISF) under grant number 1181/24.

% \section*{Supporting information}

% The following files are available free of charge.
% \begin{itemize}
%   \item Filename-1: brief description
%   \item Filename-2: brief description
% \end{itemize}

\printbibliography

\end{document}

\maketitle

\section{Static Property Predictor}

\begin{figure}[ht]
  \centering
  \includegraphics{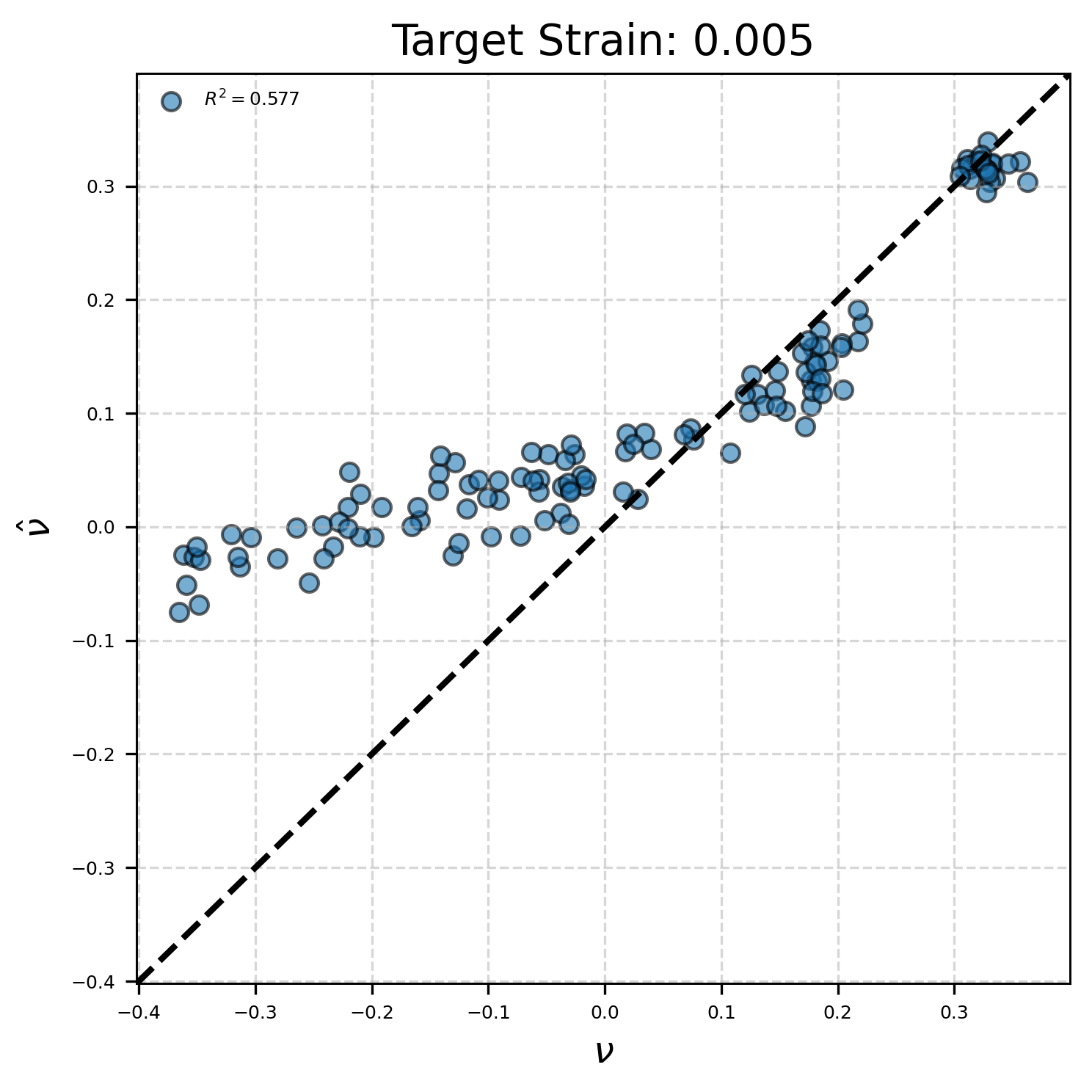}
  \caption{Performance of a static structure-property model trained on data with $\nu \geq 0.1$.}
  \label{fig::static_model_performance}
\end{figure}

\clearpage
\section{Disordered Elastic Networks and Data Generation}

\subsection{Elastic network generation}

Disordered elastic networks (DENs) were constructed in a two-step process.\cite{Rocks2017} First, $N$ soft disks of four different types and diameters were randomly placed in equal proportions within a periodic simulation box and relaxed to a local energy minimum using a standard jamming algorithm.\cite{Liu2010} The contact force between two disks $i$ and $j$, with respective radii $R_i$ and $R_j$, within the contact distance $d=R_i+R_j$ is given by
\begin{equation}
    \mathbf{F}=(k_n\delta \textbf{n}_{ij} - m_{eff}\gamma_n \textbf{v}_n) - (k_t \Delta s_t + m_{eff} \gamma_t \textbf{v}_t),
\end{equation}
where $\delta$ is the overlap distance; $\textbf{n}_{ij}$ is the unit vector along the line connecting the disk centers; $k_n$ and $k_t$ are the normal and tangential elastic constants, respectively; and $\gamma_n$ and $\gamma_t$ are the corresponding viscoelastic constants. Furthermore, $m_{eff}=\frac{M_iM_j}{M_i+M_j}$ is the effective mass of disks $i$ and $j$, whose individual masses are $M_i$ and $M_j$; $\textbf{v}_n$ and $\textbf{v}_t$ are the normal and tangential components of their relative velocity; and $\Delta s_t$ is the tangential displacement vector between the disks. Here, $k_t$, $\gamma_n$, and $\gamma_t$ were set to zero, whereas $k_n$ was set to $1.0$. The resulting force between the two disks therefore reduces to
\begin{equation}
F=\delta \textbf{n}_{ij}.
\end{equation}

During energy minimization, the system was gradually cooled from $T_{start}=5\times10^{-6}$ to $T_{end}=1\times10^{-6}$. Subsequently, a bead was placed at the center of each disk, and harmonic bonds were introduced between beads corresponding to disks in contact, $\textbf{r}_{ij}\leq d$. Each harmonic bond was assigned an elastic energy of the form
\begin{equation}\label{eq::harmonic_energy}
    E(r_{ij}) = \frac{1}{2} k_{ij} (l - l^{0})^2,
\end{equation}
where $k_{ij}$ is the harmonic bond stiffness, with units of $energy/distance^2$ (dimensionless with $k_B=1$, $\sigma = \overline{l^0}$ and $\epsilon = \overline{k_{ij}} \sigma^2$); $l^0$ is the bond rest length; and $l$ is the distance between the two beads. For each bond, $k$ was set to $1/l^0$. The resulting network was then scanned for ``dangling'' beads, defined as beads having fewer than three connections to neighboring beads. These beads were removed, and the scanning and removal procedure was repeated iteratively until none remained.

For one of the datasets, Lennard--Jones-type nonbonded interactions were additionally introduced into the elastic networks for both bonded and nonbonded beads. The interaction energy between any two beads $i$ and $j$, separated by a distance $r_{ij}$, was defined using the standard Lennard--Jones potential:
\begin{equation}
    V_{LJ}(r_{ij}) = 4\epsilon \left[ \left(\frac{\sigma}{r_{ij}}\right)^{12} - \left(\frac{\sigma}{r_{ij}}\right)^6 \right]
\end{equation}
for $r_{ij}\leq r_c$, and zero otherwise. Here, $\epsilon$ denotes the depth of the potential well, $\sigma$ is the finite separation at which the interparticle potential vanishes, and $r_c$ is the cutoff distance beyond which the interaction is truncated. Following several parameter-testing iterations, the values producing the most interesting compression dynamics were $\epsilon=0.01$, $\sigma=1.0$, and $r_c=1.122$. The latter approximates $2^{1/6}\sigma$, corresponding to the position of the potential minimum and thereby yielding an effectively repulsive interaction.

\subsection{Uniaxial compression simulation}
Compression simulations of the 2D DENs were carried out using the LAMMPS\cite{Thompson2022} molecular dynamics simulator. The systems were simulated within the NPT ensemble with anisotropic pressure control (applied to the simulation box in the direction perpendicular to compression), with constant temperature $T=1\times 10^{-10}$ , $k_B = 1$, damping parameter of $7.1 \times 10^{-3}$ and time step $\tau = 7.13 \times 10^{-6}$ controlled by the Langevin thermostat.\cite{Schneider1978} The compression was simulated using the \texttt{fix deform} command of LAMMPS with a constant engineering strain rate $s_r = 1\times10^{-8}$ and the maximum strain of 5\% along the compressed axis. Constant zero-pressure along the non-compressed axis was applied on the simulation box perpendicular to the direction of compression using the Parrinello-Rahman barostat.\cite{Parrinello1981}

\subsection{Dataset generation}

\subsubsection{Global node displacement optimization}\label{sec::opt::global_node}

Auxetic DENs were generated using a global node optimization strategy.\cite{Shen2024} This strategy is based on stochastic gradient descent and seeks to minimize Poisson's ratio, $\nu$, by modifying the positions of individual nodes. At each optimization epoch $n$, each coordinate of node $i$, denoted by $\textbf{r}_i$, is perturbed by $5\times10^{-4}$ along an axis $\eta\in{x,y}$. The bond lengths and stiffnesses are subsequently recalculated to maintain the system in its equilibrium configuration, and the gradient of the objective function is evaluated as
\begin{equation}
\frac{\partial(\nu + \mathcal{L}_1 + \mathcal{L}_2)}{\partial r_{i,\eta}},
\end{equation}
where $r_{i,\eta}$ denotes the coordinate of node $i$ along axis $\eta$, and $\mathcal{L}_1$ and $\mathcal{L}_2$ are constraints on the internode distances and bond angles, respectively. These constraints are defined as
\begin{eqnarray}
    \mathcal{L}_1 = 0.1 \sum_{i, j} H(r_\text{min} - r_{ij})(r_{ij} - r_\text{min})^2, \\
    \mathcal{L}_2 = 0.01 \sum_{j} H(\theta_\text{min} - \theta_{j})(\theta_{j} - \theta_\text{min})^6,
\end{eqnarray}
where $H$ is the Heaviside step function, while $r_{ij, \text{min}}=0.3r_{ij}^0$ and $\theta_{min}=15^{\circ}$ are empirically selected minimum values for the harmonic bond distances and bond angles, respectively. Both $\mathcal{L}_1$ and $\mathcal{L}_2$ restrict the accessible optimization space, thereby preventing the network beads from overlapping and limiting the formation of intersecting bonds. To avoid excessively large structural changes and improve the stability of the optimization, the absolute value of each gradient component was clipped at $0.01$. Finally, the node positions were updated using the $N\times2$ gradient matrix obtained in the preceding step:
\begin{equation}
    r_{i,\eta}^{n+1} = r_{i,\eta}^{n} - \lambda \frac{\partial(\nu + \mathcal{L}_1 + \mathcal{L}_2)}{\partial r_{i,\eta}},
\end{equation}
where $\lambda$ is the learning rate. This procedure was repeated until either \textit{(a)} the desired target value of $\nu$ was reached or \textit{(b)} the number of iterations exceeded a predefined maximum.

\subsubsection{Harmonic bond stiffness tuning}\label{sec::opt::stiff}

Using the minimal differentiable PyTorch-based MD engine described in Section \ref{sec::torch-sim}, we investigated two distinct optimization schemes, referred to as \textbf{continuous} and \textbf{discrete} optimization. In both schemes, the learnable parameters are adjustments to the individual bond stiffnesses $k_{ij}$. These adjustments are denoted by $\Delta k_{ij}$, where the directed graph edge represents a harmonic bond between the two network beads $i$ and $j$.

In the \textbf{continuous optimization} scheme, each initial harmonic bond stiffness $k_{ij}^0$ is adjusted continuously according to
\begin{equation}
    k_{ij}^{\prime} = k_{ij}^0 \left(1 + C \cdot \tanh(\Delta k_{ij})\right),
\end{equation}
where $C$ is a predefined constant that limits the maximum relative change in stiffness.

Conversely, the \textbf{discrete optimization} scheme, inspired by a procedure reported elsewhere,\cite{Reid2018} restricts the allowed changes in bond stiffness in a manner resembling discrete bond pruning. This is achieved using an edge mask $g_{ij}$:
\begin{eqnarray}
    g_{ij} &=& \sigma(\beta \Delta k_{ij}^\text{sym}), \\
    k_{ij}^{\prime} &=& k_{ij}^0 g_{ij},
\end{eqnarray}
where $\sigma$ is the sigmoid function and $\beta$ is a hyperparameter controlling its steepness. Due to the undirected nature of the graph, the before each update the symmetric stiffness adjustment $\Delta k_{ij}^\text{sym}$ is computed:
\begin{equation}
    \Delta k_{ij}^\text{sym} = \frac{\Delta k_{ij} + \Delta k_{ji}}{2}
\end{equation}
At each optimization epoch, a DEN with the updated stiffnesses $k^{\prime}$ is compressed for a fixed number of MD steps. The Poisson's ratio $\nu$ of the network is calculated directly from the changes in the periodic box dimensions between the initial ($t=0$) and final ($t$) simulation steps:
\begin{equation}
    \nu = -\frac{(L_y^t - L_y^0)/L_y^0}{(L_x^t - L_x^0)/L_x^0}.
\end{equation}
The optimization objective is to minimize the mean squared error (MSE) between the measured Poisson's ratio and its target value. The total loss function $\mathcal{L}$ is defined as
\begin{equation}
\mathcal{L} = (\nu - \nu_\text{target})^2 + \lambda \frac{1}{M} \sum_{ij} (1 - g_{ij}),
\end{equation}
where $M$ is the total number of unique edges. The second term is a bond removal penalty term used exclusively in the discrete mode, and $\lambda$ is its corresponding weight. For the continuous mode, $\lambda$ is set to zero.

Optimization was performed using stochastic gradient descent (SGD) with momentum. Before each optimizer step, the gradients with respect to the stiffness-adjustment parameters $\Delta k_{ij}$ were clipped to a maximum norm of $1.0$. The procedure was continued for a predefined number of epochs or terminated early if the simulated network achieved the target Poisson's ratio $\nu_\text{target}$.

\section{GNN MD Simulator}\label{sec::gnn-simulator}

\subsection{Model architecture}

The model adopts an Encoder--Processor--Decoder architecture.\cite{Hamrick2018RelationalMachines, Pfaff2021} The encoder maps the input node and edge features of a graph $G$ to latent representations. The temporal history of the node features ${\textbf{x}_i}$ is first separated into its spatial components ($x$ and $y$). Each component is processed using the same axis-shared MLP before the resulting representations are flattened and linearly projected onto a hidden space of dimension 128. The edge features ${\textbf{e}_i}$ are mapped directly to the 128-dimensional latent space using an independent MLP, denoted by $\varepsilon^e$.

The processor consists of $k$ message-passing (MP) layers with identical architectures but independent sets of learnable parameters. These layers are applied sequentially, with each layer operating on the output of the preceding layer and incorporating a residual connection. The node embeddings are updated according to
\begin{equation}
    \textbf{x}_{i}^{(k)} = \textbf{x}_{i}^{(k-1)} + \psi^{(k)} \left( \textbf{x}_{i}^{(k-1)}, \sum_{j\in\mathcal{N}(i)}\phi^{(k)}(\textbf{x}_{i}^{(k-1)}, \textbf{x}_{j}^{(k-1)},\textbf{e}_{ij}) \right),
\end{equation}
where $\psi$ and $\phi$ are learnable MLPs employing layer normalization. Here, $\textbf{x}_{i}^{(k-1)}$ denotes the embedding of node $i$ in the $(k-1)$th processor layer, and the summation is taken over all nodes $j$ bonded to node $i$.

The processor output is mapped by the decoder to a two-dimensional output ${\hat{\textbf{a}}_i} \in \mathbb{R}^{N \times 2}$ representing the predicted node accelerations. The decoder is implemented using an additional MLP denoted by $\delta^x$. All MLPs in the architecture, including $\varepsilon^x$, $\varepsilon^e$, $\psi$, $\phi$, and $\delta^x$, were implemented as feed-forward networks with hidden dimensions of 128 and ReLU activation functions.

\subsection{Implementation and training details}\label{sec::gnn_simulator::training}

The model was implemented using PyTorch Geometric 2.7.0\cite{Fey2019, Fey2025PyGGraphs} and PyTorch 2.11.0.\cite{Paszke2019, Ansel2024PyTorchCompilation} The datasets were generated from uniaxial compression trajectories of DENs spanning a range of Poisson's ratios $\nu \in [-0.6,0.6]$.

The input data, comprising the node and edge features, and the output data were normalized online and independently during training to have zero mean and unit variance. Specifically, at each forward step, the data were normalized using statistics accumulated from all values observed by the model up to that point.\cite{Sanchez-Gonzalez2020, Kumar2022GNS:Modeling} To stabilize the later stages of training, the normalizer statistics were frozen after a predefined epoch, typically epoch 5. All learnable weights and biases were randomly initialized from a normal distribution with zero mean and a standard deviation of $1/\sqrt{n}$, where $n$ is the number of weights in the corresponding layer. We found that initialization using $\mathcal{N}(0,1/\sqrt{n})$ generally produced better performance than initialization using the uniform distribution $\mathcal{U}(-1/\sqrt{n},1/\sqrt{n})$.

Optimization was performed using the Adam optimizer with zero weight decay. The learning rate was reduced after each epoch using an exponential scheduler with $\gamma=0.995$. Gradient clipping with a maximum norm of $1.0$ was applied together with gradient accumulation. The model was trained in a supervised manner by minimizing the Huber loss\cite{Huber1992RobustParameter} between the predicted accelerations $\hat{\textbf{a}}$ and the normalized ground-truth accelerations $\textbf{a}$:
\begin{equation}
L_{\delta}(\textbf{a}, \hat{\textbf{a}}) =
    \begin{cases}
        \frac{1}{2}(\textbf{a} - \hat{\textbf{a}})^2 & \text{if } |\textbf{a} - \hat{\textbf{a}}| \le \delta \\
        \delta |\textbf{a} - \hat{\textbf{a}}| - \frac{1}{2}\delta^2 & \text{otherwise}
    \end{cases}
\end{equation}
where $\delta$ is the threshold separating the quadratic and linear regimes of the loss.

Training was performed using a multi-step supervision (MST) strategy.\cite{Shteingolts2026EnablingSimulators} For each training sample during the forward pass the simulator model performs a short, continuous rollout. To maintain the training stability, the rollout length increases as a function of training epochs, starting at a single step and scaling up to 10 steps in later epochs. The autoregressive training loop is dynamically coupled with the barostat algorithm.\cite{Shteingolts2026EnablingSimulators} At each rollout step, the simulation box dimensions $L_x$ and $L_y$ are updated to maintain the target pressure in the transverse direction $P_{yy} = 0.0$. Edge attributes and per-particle forces are iteratively recomputed based on the evolving box dimensions before being fed back into the model for the subsequent prediction step.

\section{Simulator-coupled barostat}\label{sec::barostat}

\subsection{Implementation}

At each timestep $t$, the periodic box algorithm computes the instantaneous transverse virial $P_y^V$ and kinetic $P_y^K$ pressure from the predicted node positions and pairwise forces. The kinetic term is computed using the ideal gas approximation as follows:
\begin{equation}
    P_y^K = \frac{1}{2}\frac{k_B NT}{A},
\end{equation}
The virial contribution is calculated by summing the pairwise force contributions from all active interaction edges:
\begin{equation}
    P_y^V = \frac{1}{2A} \sum f_{ij,y}r_{ij,y},
\end{equation}
where $A=L_xL_y$ is the box area and $f_{ij,y}$ and $r_{ij,y}$ are the $y$-components of the pairwise force and minimum-image separation vector, respectively. The factor of $1/2$ corrects for the bidirectional representation of each physical interaction.

For a harmonic bond, the pairwise force is
\begin{equation}
    \mathbf{f}_{ij}^\text{harm} = -k_{ij}(r_{ij}-r_{ij}^0),
\end{equation}
where $k_{ij}$ is the bond stiffness, $r_{ij}=|\mathbf{r}_{ij}|$ is the current bond length, and $r_{ij}^0$ is its rest length. When nonbonded Lennard--Jones interactions are present, the force associated with an LJ interaction edge is
\begin{equation}
    \mathbf{f}_{ij}^\text{LJ} = \frac{24\epsilon}{r_{ij}}
    \left[
        2\left(\frac{\sigma}{r_{ij}}\right)^{12} -
        \left(\frac{\sigma}{r_{ij}}\right)^6
    \right],
\end{equation}
and is zero for $r_{ij}\geq r_c$. Here, $\epsilon$ and $\sigma$ are the Lennard--Jones parameters and $r_c$ is the interaction cutoff. The force used in the virial calculation is therefore computed according to the present interaction types. For networks without nonbonded interactions the virial contains only harmonic contributions.

The difference between the internal pressure $P_y$ and the target pressure $P_t$ generates a thermodynamic driving force $F_d$. We define a total force $F_\text{total}$ and apply a frictional damping coefficient $\gamma$ to prevent aggressive oscillations:
\begin{eqnarray}
    F_{d} = (P_y - P_t) \cdot L_x \\
    F_\text{total} = F_d - \gamma \cdot v^t_y
\end{eqnarray}
where $v^t_y$ is the box velocity along the $y$ axis. Box acceleration $a^t_y$ can then be computed using the piston mass $W_y$:
\begin{equation}
    \textbf{a}^t_y = \frac{F_\text{total}}{W_y}
\end{equation}
Finally, the box velocity is updated and the next-step $L_y$ is calculated as follows:
\begin{eqnarray}
    v^{t+1}_y = v^{t}_y + a^t_y \cdot \Delta t \\
    L^{t+1}_y = L_y^t + v^{t+1}_y\Delta t
\end{eqnarray}

\subsection{Empirical parameter tuning}

Because the periodic box adjustment procedure operates on coarse-grained dynamics (a single GNN step corresponds to 200 MD timesteps), the physical constants typically used for piston mass $W_y$ and damping coefficient $\gamma$ in classical barostats are not directly applicable. Consequently, the specific values of both $\gamma$ and $W_y$ must be empirically tuned. 

To this end, we utilize the Optuna hyperparameter optimization framework. \cite{Akiba2019Optuna} The piston mass and damping coefficient are defined as functions of the number of particles $N$ and the coarse-grained time stride $\Delta t$:
\begin{eqnarray}
    W_y = C_\text{coupling} \cdot N \cdot (\Delta t)^2 \\
    \gamma = C_\text{damping} \cdot N \cdot \Delta t
\end{eqnarray}
where $C_\text{coupling}$ and $C_\text{damping}$ are the dimensionless scalar factors to be optimized. 

To find the optimal values for $C_\text{coupling}$ and $C_\text{damping}$, we minimize a cumulative objective function. The objective function calculates the discrepancy between the predicted transverse box dimension $L_y^\text{pred}$ and the ground-truth dimension $L_y^\text{GT}$ using a Huber loss. During the optimization procedure, we utilize ground-truth particle positions at each step to isolate the box adjustment from GNN prediction errors. In our testing, 200 optimization trials were enough to fully converge on optimal values. The search space for $C_\text{coupling}$ was defined as $[1.0, 15.0]$ and for $C_\text{damping}$ as $[0.5, 10.0]$. The resulting best parameters were then fixed for all subsequent rollout experiments and MST training regimes.

\section{Minimal differentiable MD simulator}\label{sec::torch-sim}

We implemented a compact molecular dynamics (MD) engine natively in PyTorch. The simulator performs uniaxial compression along the $x$-axis while allowing the transverse box dimension $L_y$ to fluctuate dynamically in response to the internal pressure. The simulation is differentiable with respect to continuous model parameters for a fixed interaction graph and realization of the stochastic noise.

Compression along the $x$-direction is imposed deterministically at a constant engineering strain rate $\dot{\epsilon}$. At simulation step $n$, corresponding to the physical time $t_n=n\Delta t$, the longitudinal box dimension is defined as
\begin{equation}
    L_x(t_n) = L_x(0)(1-\dot{\epsilon}t_n).
\end{equation}
Particle coordinates along the $x$-axis are affinely rescaled by the ratio of the new and previous box dimensions. In addition to this affine deformation, the particle positions are propagated using their half-step velocities, thereby allowing nonaffine motion along both spatial directions.

The transverse box dimension is coupled to a dynamic barostat that seeks to maintain a target pressure $P_\text{target}$. The instantaneous transverse pressure is evaluated as
\begin{equation}
    P_{yy} = \frac{1}{V}
    \left(
        \sum_{i=1}^{N}m v_{i,y}^{2}
        +\frac{1}{2}\sum_{(i,j)\in\mathcal{E}_\text{dir}}
        F_{ij,y}r_{ij,y}
    \right),
\end{equation}
where $V=hL_xL_y$ is the effective simulation volume, with a fixed out-of-plane thickness $h=0.2$. The second sum is evaluated over the directed interaction edges $\mathcal{E}_\text{dir}$. The factor of $1/2$ removes the double counting that arises when each physical interaction is represented by two oppositely directed edges. Here, $m$ is the particle mass, $v_{i,y}$ is the transverse particle velocity, and $F_{ij,y}$ and $r_{ij,y}$ are the $y$-components of the pairwise force and minimum-image separation vectors, respectively.

The barostat variable $v_{b,y}$ represents the logarithmic strain rate of the transverse box rather than the ordinary velocity of its boundary. Its acceleration is defined as:
\begin{equation}
    a_{b,y} =
    \frac{\left(P_{yy}-P_\text{target}\right)V}{W_y},
\end{equation}
where the piston inertia is defined as
\begin{equation}
    W_y=(2N-3)k_\text{B}T_b\tau_b^2.
\end{equation}
Here, $\alpha$ is an inertia prefactor, $T_b$ is the barostat temperature, and $\tau_b$ is the barostat damping time. The piston degree of freedom is coupled to a three-variable Nosé--Hoover chain with mass parameter $Q_b=k_\text{B}T_b\tau_b^2$. The barostat and thermostat-chain variables are integrated using five subcycles during each half-step. The accumulated barostat scaling is applied to the transverse particle velocities, and the box dimension is propagated according to
\begin{equation}
    L_y(t+\Delta t)
    =
    L_y(t)\exp\left(v_{b,y}^{\,1/2}\Delta t\right),
\end{equation}
where $v_{b,y}^{\,1/2}$ denotes the barostat rate obtained after the predictor half-step. The particle equations of motion are integrated using a stochastic velocity-Verlet-like scheme with a default timestep $\Delta t=0.01$.

A complete integration step proceeds as follows:
\begin{enumerate}
    \item The deterministic forces and virial pressure are evaluated using the current particle coordinates, box dimensions, and interaction graph.
    \item The barostat velocity and its three Nosé--Hoover-chain variables are advanced through five predictor subcycles using the current pressure difference. The resulting metric scaling factor is accumulated over these subcycles.
    \item A Gaussian random force is sampled, and the particle velocities are advanced by $\Delta t/2$ using the deterministic forces, Langevin drag, stochastic force, and transverse metric scaling.
    \item The new longitudinal box length is evaluated from the prescribed engineering-strain schedule, while the transverse box length is propagated exponentially using the half-step barostat rate. Particle coordinates in both directions are updated by combining affine box rescaling with displacement according to the half-step velocities.
    \item The deterministic forces and transverse pressure are recalculated using the updated coordinates and box dimensions.
    \item The barostat and Nosé--Hoover-chain variables are advanced through five corrector subcycles using the updated pressure.
    \item Particle velocities are advanced through the final $\Delta t/2$ using the updated deterministic forces and Langevin drag, together with the same random-force realization used in the predictor step.
\end{enumerate}

After each full integration step, the graph representation is refreshed. Bond stiffnesses are retained, whereas distance-dependent edge attributes and, when applicable, nonbonded interaction edges are updated. Every 100 MD steps, both the center-of-mass translational velocity and the net rigid-body rotational velocity are removed.

\clearpage
\section{Inference-Time Physics-based Optimization}

\subsection{Physics constraints}

The net force on each particle $\mathbf{F}_i$ is derived from the harmonic bond stiffness $k_{ij} = 1/l^0_{ij}$:
\begin{equation}
    \mathbf{F}_i = \sum_{j \in \mathcal{N}(i)} -k_{ij} (l_{ij} - l^0_{ij}) \frac{\mathbf{r_{ij}}}{l},
\end{equation}
where $l_{ij}$ and $l^0_{ij}$ are the current and rest bond lengths between the beads $i$ and $j$, while $\mathbf{r_{ij}}$ is the bond vector. When nonbonded Lennard--Jones interactions are present, the total force is the sum of the harmonic and nonbonded contributions: 
\begin{equation} 
    \mathbf{F}_i = \mathbf{F}_i^\text{harm} + \mathbf{F}_i^\text{LJ}
\end{equation}
The Lennard--Jones contribution is given by
\begin{equation}
\mathbf{F}_i^\text{LJ} = \sum_{j\in\mathcal{N}_\text{LJ}(i)} \frac{24\epsilon}{r_{ij}} \left[ 2\left(\frac{\sigma}{r_{ij}}\right)^{12} - \left(\frac{\sigma}{r_{ij}}\right)^6 \right]
\end{equation}
and is zero for $r_{ij}\geq r_c$. Here, $r_{ij}=|\mathbf{r}_{ij}|$, $\epsilon$ and $\sigma$ are the Lennard--Jones parameters, $r_c$ is the interaction cutoff, and $\mathcal{N}_\text{LJ}(i)$ denotes the set of nonbonded neighbors of particle $i$ within the cutoff $r_c$.

The force loss component penalizes any non-zero forces along the transverse axis $y$ across the $N$ particles and is defined as:
\begin{equation}
    \mathcal{L}_\text{force} = \frac{1}{N} \sum (F_{i, y})^2
\end{equation}
The potential energy constraint $\mathcal{L}_U$ is defined as a sum of individual harmonic bond energies $E(\textbf{r}_{ij})$ (Equation \ref{eq::harmonic_energy}):
\begin{equation}
    \mathcal{L}_U = \sum E(\textbf{r}_{ij})
\end{equation}
The pressure loss term is defined as the squared virial stress:
\begin{equation}
    \mathcal{L}_\text{pressure} = (\sigma_{yy})^2
\end{equation}
where $\sigma_{yy}$ is computed using the pairwise harmonic forces $\mathbf{f}_{ij, y}$ and edge vectors $\mathbf{r}_{ij, y}$ over area $A$:
\begin{equation}
   \sigma_{yy} = \frac{1}{2A} \sum f_{ij, y} \; r_{ij, y} 
\end{equation}

\subsection{Empirical weights tuning}

Maximizing ITPO performance requires carefully tuning the individual weights $\alpha_i$ of the physics constraints:
\begin{equation}
    \mathcal{L}_\text{total} = \mathcal{L}_\text{anchor} + \sum_i \alpha_i \mathcal{L}_i^\text{phys},
\end{equation}
This presents a practical challenge. Because our objective is OOD generalization to unseen dynamics (typically to $\nu < 0.1$, while the training data is characterized by $\nu \geq 0.1$), ITPO weights cannot be found using these data. However, if the GMDS has already been trained using the MST strategy on data extending down to $\nu = 0.1$, it can sometimes perform \textit{too well}. Because of this, the hyperparameter search converges on virtually zero physics weights $\alpha_i \approx 0$, as the GMDS predictions are already plenty accurate. To ensure the physics constraints remain active, we utilize a staggered training and tuning pipeline:
\begin{enumerate}
    \item First, a GMDS model is trained strictly on \textit{in-distribution} data characterized by $\nu > 0.2$ using the MST strategy.
    \item Next, a search for ITPO physics constraint weights $\alpha_i$ is performed using the intermediate finetuning data with $\nu \in [0.1, 0.2]$. Because the GMDS is not familiar with this dynamics, optimizer relies on the physics constraints to minimize the ITPO objective function, yielding non-zero constraint weights $\alpha_i$.
    \item Finally, the model used for the search is further trained on the intermediate data with $\nu \in [0.1, 0.2]$ using a MST strategy.
\end{enumerate}

To find individual physics constraint weights $\alpha_i$ we employ the Optuna hyperparameter optimization framework.\cite{Akiba2019Optuna} The objective function is defined through the coefficient of determination ($R^2$) between the predicted and ground-truth Poisson's ratios $\nu$ over a 50-step rollout. In total, up to 5 different parameters can be optimized:
\begin{enumerate}
    \item Refinement iterations $N_\text{iter}$: The number of gradient descent steps performed at each rollout timestep, sampled uniformly in the range $[5, 40]$.
    \item Learning rate $\eta$ for the Adam optimizer, sampled on a log-uniform scale between $10^{-8}$ and $10^{-3}$.
    \item The scalar coefficients for the force $\alpha_\text{force}$, potential energy $\alpha_\text{energy}$, and virial pressure $\alpha_\text{pressure}$ terms. These are sampled log-uniformly between $10^{-7}$ and $10^{-1}$ to account for the varying orders of magnitude of the underlying physical quantities.
\end{enumerate}
In practice, we found it beneficial to leave the number of refinement iterations $N_\text{iter}$ constant, since it plays a similar role to the learning rate $\eta$. Depending on the dataset, $N_\text{iter}$ was set to 20 or 30, which presents a compromise between the refinement accuracy and the computational penalty incurred due to ITPO. Overall, 500 Optuna optimization trials were performed per dataset.

\subsection{Gradient computation via the Implicit Function Theorem}

A significant challenge for the proposed end-to-end differentiable inverse design framework is backpropagating the macroscopic property gradients through the unrolled inner ITPO loop. Directly tracking gradients through inner optimization steps at every rollout timestep is computationally intractable. To circumvent this, we utilize the Implicit Function Theorem (IFT) to analytically compute the approximate implicit gradients at the local minimum of the ITPO loss function.

Assuming the ITPO reaches a stable minimum at $\mathbf{a}_\text{nn}^*$, the gradient of the total loss with respect to the initial GNN prediction $\hat{\mathbf{a}}_\text{nn}$ can be evaluated implicitly. Let $H = \nabla^2_{\mathbf{a}_\text{nn}^*} \mathcal{L}_\text{total}$ represent the Hessian matrix of the loss evaluated at the refined state. Applying IFT requires calculating the product of the inverse Hessian and the upstream gradient vector, $\mathbf{v}$. We employ a matrix-free Hessian-vector product (HVP):
\begin{equation}
    \text{HVP}(\mathbf{v}) = \nabla_{\mathbf{a}_\text{nn}^*} \left( \left[ \nabla_{\mathbf{a}_\text{nn}^*} \mathcal{L}_\text{total} \right] \cdot \mathbf{v} \right)    
\end{equation}
To solve the linear system $H \mathbf{x} = \mathbf{v}$ for the inverse HVP, we utilize a conjugate gradient solver. Finally, the implicit gradient must account for the mixed partial derivative between the refined and initial accelerations. Given that the anchor loss is defined as a mean squared error:
\begin{equation}
    \mathcal{L}_\text{anchor} = \frac{1}{N} || \mathbf{a}_\text{nn}^* - \hat{\mathbf{a}}_\text{nn} ||^2    
\end{equation}
where $N$ is the total number of particles, the mixed derivative simplifies to a scaled identity operator. Consequently, the analytical gradient mapped back to the simulator model output is computed as:
\begin{equation}
    \nabla_{\hat{\mathbf{a}}_\text{nn}} \mathcal{L} = \frac{2}{N} \left( H^{-1} \mathbf{v} \right)
\end{equation}

% \clearpage
\printbibliography